\documentclass[%
A4,
prx,
aps,
amsmath,
amssymb,
nopacs,
twocolumn,
]{revtex4-1}

\usepackage{amsmath}
\usepackage{amssymb}
\usepackage{graphicx}
\usepackage{amsthm}

\usepackage{latexsym}
\usepackage[dvipsnames]{xcolor}
\usepackage[colorlinks=true, urlcolor=BlueViolet, linkcolor=BlueViolet, citecolor=BlueViolet]{hyperref}
\usepackage{etoolbox}

\newcommand{\dd}{\mathrm{d}}
\newcommand{\Eexp}[1]{ \ensuremath{ \textrm{e}^{ #1} } }
\newcommand{\eqText}[1]{ \quad \text{#1} \quad }
\newcommand{\BoldVec}[1]{ \ensuremath{ {\boldsymbol{#1}} } }

\DeclareMathOperator{\tanhc}{tanhc}

\newcommand{\Creation}[2]{ \ensuremath{ #1_{#2}^{\dagger} } }
\newcommand{\Annihilation}[2]{ \ensuremath{ #1_{#2} } }
\newcommand{\NN}[2]{ \ensuremath{ {\left\langle #1,#2 \right\rangle} } }
\newcommand{\NumberOp}[2]{ \ensuremath{ { #1 _{#2}^{\dagger} #1_{#2} } }}

\newcommand{\Abs}[1]{ \ensuremath{ { \left| #1 \right| } } }
\newcommand{\Hopping}[4]{ \ensuremath{ {#1} _{#2}^{\dagger} {#3} _{#4} } }
\newcommand{\BCSTermUp}[4]{\ensuremath{ { #1 _{#2 \uparrow }^{\dagger} #3 _{#4 \downarrow }^{\dagger}} }}

\newcommand{\PP}{\ensuremath{^{+} } }
\newcommand{\PH}{\ensuremath{^{-} } }
\newcommand{\PPorPH}{\ensuremath{^{\pm} } }

\begin{document}

\title{Universal phase diagrams with superconducting domes for electronic flat bands}
\author{Tomas L\"{o}thman}
\author{Annica M. Black-Schaffer}
\affiliation{Department of Physics and Astronomy, Uppsala University, Box 516, S-751 20 Uppsala, Sweden}
\date{\today}



\begin{abstract}	
	Condensed matter systems with flat bands close to the Fermi level generally exhibit, due to their very large density of states, extraordinary high critical ordering temperatures of symmetry breaking orders, such as superconductivity and magnetism. Here we show that the critical temperatures follow one of two universal curves with doping away from a flat band depending on the ordering channel, which completely dictates both the general order competition and the phase diagram. Notably, we find that orders in the particle-particle channel (superconducting orders) survive decisively further than orders in the particle-hole channel (magnetic or charge orders) because the channels have fundamentally different polarizabilities. Thus, even if a magnetic or charge order initially dominates, superconducting domes are still likely to exist on the flanks of flat bands. We apply these general results to both the topological surface flat bands of rhombohedral ABC-stacked graphite and to the van Hove singularity of graphene.
	
\end{abstract}

\maketitle

\section{INTRODUCTION}
	From the high-temperature cuprates \cite{HTCRevModPhys.66.763,HTCRevModPhys.72.969} to the families of iron-based \cite{paglione2010high,RevModPhys.87.855} and heavy fermion \cite{HFRevModPhys.81.1551} superconductors, many materials of significant contemporary interest have complex phase diagrams with neighboring (anti) ferromagnetic and superconducting states that can be tuned from one to the other by e.g.~doping. Symmetry breaking orders, such as these, emerge at a critical temperature, at which the interactions favoring ordering overcome the thermal disorder. Stronger interactions therefore give a higher critical temperature, but so does a larger density of states (DOS) near the Fermi level.
	
	Energy bands with a low dispersion have large DOS, culminating in a divergent DOS for flat bands. While previously thought to be uncommon outside Landau levels and some special lattice structures 
	\cite{PhysRevLett.114.245504,FBLattices}, 
	flat bands have recently been found to also exist as protected boundary states of topologically non-trivial electronic structures 
	\cite{PhysRevB.84.235126,PBSingapless,PhysRevLett.89.077002,Heikkil2011,FBTM2}. 
	For example, topology ensures that the nodal Dirac cones of graphene are connected by a flat band on the zigzag edges of graphene ribbons \cite{PhysRevB.73.235411,Fujita_1996,PhysRevB.54.17954,PhysRevB.73.125415,PhysRevB.73.085421,PhysRevB.71.193406}. Similarly, approximately flat surface states, called drumhead states, have recently been found for topological line-node semimetals
	\cite{DrumheadFBPhysRevB.92.045108,DrumHeadPhysRevLett.115.036807,DrumHeadPhysRevB.93.205132,DrumHeadPhysRevB.93.235147},
	nodal-chain metals 
	\cite{Bzduek2016}, 
	and topological nodal superconductors have been shown to host Majorana surface flat bands 
	\cite{PhysRevB.88.060504}.
	
	A large topologically protected surface flat band has recently also been found for the line-node semimetal rhombohedral, or ABC-stacked, graphite 
	\cite{PhysRevB.73.245426,pierucci2015evidence,Heikkil2011,PhysRevB.83.220503,henni2016rhombohedral}. Here density-functional theory (DFT) calculations have shown a strong ferrimagnetic ordering in the surface flat band states
	\cite{PhysRevB.81.161403,Cuong2012253}.
	But ABC-stacked graphite has recently also been evoked as a candidate for high-temperature superconductivity 
	\cite{FBTM2,PhysRevB.87.140503,PhysRevB.83.220503}, since both its superconducting critical temperature and supercurrent have been shown to increase linearly with the interaction strength and the area of the surface flat band
	\cite{PhysRevB.83.220503,Kopnin2011,peotta2015superfluidity}.		
	Such a linear relationship for the superconducting critical temperature has also been found in the flat pseudo-Landau-levels of strained graphene \cite{PhysRevLett.111.046604}. In fact, many of the recently discovered flat bands systems show an enhanced susceptibility towards superconductivity \cite{PhysRevB.83.220503,PhysRevB.87.140503}. However, alternative orders have  also been shown to be strongly enhanced, including flat band ferromagnetism \cite{Tasaki_1998, PhysRevLett.62.1201} and robust magnetic order along the zigzag edge of graphene \cite{PhysRevLett.100.047209,Bhowmick_2008,PhysRevB.79.235433,PhysRevLett.106.226401,Li_2016,Magda_2014}. Thus, while ordering is very often expected in flat band systems, it is not generally known if the large DOS peak actually favors superconductivity or other orders. Moreover, very little attention has been given to the competition between different orders and consequently no general phase diagram for flat band systems has been developed.

In this work we establish a universal phase diagram for flat band systems, including all possible superconducting particle-particle (PP) and magnetic and/or charge particle-hole channel (PH) orders. 
More specifically, we first show that all symmetry breaking orders in any flat band system show a similar enhancement because of the large DOS and exhibit a linear scaling of their critical temperatures with the interaction strength. 
We then establish that the critical temperatures of PP and PH orders follow their own unique universal expressions as a function of the doping away from the flat band for any set of interactions. In fact, we find that all the details of the interactions are possible to fully encapsulate in the critical temperatures found when the Fermi level coincides with the flat band energy ($\hat{T}_c\PP$ for PP and $\hat{T}_c\PH$ for PH), and therefore only $\hat{T}_c\PPorPH$ and the doping level enter the final expressions. 
With these particularly simple expressions we develop the completely general phase diagram of any flat band system, which also directly addresses order competition. 
Surprisingly, we find that superconducting domes very likely appear on the flanks of flat bands. In fact, even if a PH order is formed substantially before a competing superconducting order upon cooling when the Fermi energy is aligned with the flat band, a superconducting dome will always appear on both sides of the flat band as long as $\hat{T}_c\PP \gtrsim \frac{1}{2}\hat{T}_c\PH$.

	We are able to establish the general phase diagram exactly for any interactions in an ideally flat band system. Remarkably, we also show that the results remain valid for all DOS peaks that are narrow compared to the energy scale of the interactions. We are therefore able to apply our results not only in ideal flat bands systems, but also for the approximate flat surface bands, such as those found in finite ABC-stacks of graphite and at the van Hove singularity (VHS) in heavily doped graphene \cite{RevModPhys.81.109}. 
	For ABC-stacked graphite we show that even if a magnetic order is initially found on the surface, a superconducting state can still be accessible by either  doping or applying an electric field across the graphite stack. 
	Moreover, we find that the locally flat band saddle points of the VHS in heavily doped graphene fully dictate the general order competition near the VHS. This includes the characteristic superconducting dome structure of our flat band results. Interestingly, this offers a clear explanation to the recent (functional) renormalization group (f)RG results that have all found a spin-density-wave (SDW) at the VHS, but chiral $d$-wave superconducting domes on both flanks of the VHS \cite{PhysRevB.86.020507,PhysRevB.85.035414}.

\section{CRITICAL TEMPERATURES}
	Our main results are derived from the structure of the critical temperature equations for general symmetry breaking ordered states. We start our treatment from a completely general translationally invariant Hamiltonian $H$ with a spin and particle number conserving quadratic part $H_0$ expressed through energy bands $\xi_\alpha \! \left( \BoldVec{k} \right)$ and a set of general two-body interaction potentials $V$ in $H_{\rm int}$:
		\begin{align} \label{eq:H}
			H 
			&
			= 
			H_0 + H_\text{int}
			\nonumber \\
			&
			=
			\sum_{\BoldVec{k} \sigma \alpha}
			\xi_\alpha \! \left( \BoldVec{k} \right)
			\Creation{c}{\BoldVec{k} \sigma \alpha}
			\Annihilation{c}{\BoldVec{k} \sigma \alpha}
			\\
			&
			+
			\sum_{
				{
					\BoldVec{k} \BoldVec{p} \BoldVec{q}	
					\alpha \beta \gamma \delta					
					1 2 3 4																
				}
			}
			V^{1234}_{\alpha \beta \gamma \delta} \!
			\left(
				\BoldVec{k}, \BoldVec{p} , \BoldVec{q}
			\right)
			\Creation{c}{
				\BoldVec{k} 1 \alpha
			}
			\Creation{c}{
					\BoldVec{p}	2 \beta 
			}
			\Annihilation{c}{
					\BoldVec{p}+\BoldVec{q} 3 \gamma 
			}
			\Annihilation{c}{
					\BoldVec{k}-\BoldVec{q} 4 \delta 
			}
			\nonumber 
			,
	\end{align}
	where Greek indices label electronic bands, numbers label spins, and $\BoldVec{k}$, $\BoldVec{p}$, and $\BoldVec{q}$ label crystal momenta. To capture all conventional ordered states, we decouple $H_{\rm int}$ completely into mean-field order parameters and omit a constant energy shift:
	\begin{align} \label{eq:HMF}
		H_{\text{MF}}
		&
		=
		\sum_{
			{
				\BoldVec{k} \BoldVec{p}	
				\alpha \beta 						
				12															
			}
		}
		\left[
			d_{\alpha \beta} \! 
			\left( 
				\BoldVec{k},\BoldVec{p}
			\right)
			\cdot
			\chi
		\right]_{12}
		\Creation{c}{
			\BoldVec{k} 1 \alpha
		}
		\Creation{c}{
				\BoldVec{p} 2 \beta 
		}
		+
		\text{H.c.}
		\nonumber \\
		&
		\quad
		+
		4
		\sum_{
				\BoldVec{k} \BoldVec{q}
				\alpha \delta						
				14															
		}
		\left[
			g_{\alpha \delta} \! 
			\left( 
				\BoldVec{k},\BoldVec{q}
			\right)
			\cdot
			\sigma
		\right]_{14}
		\Creation{c}{
			\BoldVec{k} 1 \alpha
		}
		\Annihilation{c}{
				\BoldVec{k}-\BoldVec{q} 4 \delta 
		}
		.
	\end{align}
	There are two types of mean-field order parameters. There are the superconducting orders in the PP channel with order parameters $d_{\alpha \beta} \! \left( \BoldVec{k},\BoldVec{p} \right)$. These correspond to pairing between the momentum states at $\BoldVec{k}$ and $\BoldVec{p}$, where a non-zero total momentum $\BoldVec{k} + \BoldVec{p}$ is characteristic of an FFLO order \cite{PhysRev.135.A550}. There are also magnetic and charge orders in the PH channels with order parameters $g_{\alpha \delta} \! \left( \BoldVec{k},\BoldVec{q} \right)$, where $\BoldVec{q}$ is the spatial modulation wave vector.
	The only approximation introduced here is the omission of the interactions between the fluctuations away from the constant order parameter values, which in well-ordered states are small. Moreover, the BCS wave function implicit in the above PP channel decoupling has been shown to be an exact ground state of several flat band systems and to accurately capture the properties of their superconducting state \cite{PhysRevLett.117.045303,peotta2015superfluidity}, which further supports a mean-field approach. The order parameters are defined self-consistently by demanding that the quadratic Hamiltonian $H_0 + H_{\rm MF}$ minimizes the free energy,	
	\begin{align} \label{eq:DandG}
			d^{\mu}_{\alpha \beta} \! 
			\left( 
				\BoldVec{k},\BoldVec{p}
			\right)
			&
			=
			\frac{1}{2} 
			\sum_{
				{
					\BoldVec{q}							
					\gamma \delta						
					1 2 3 4															
				}
			}
			[ \chi^\mu ]^\dagger_{21}
			V^{1234}_{\alpha \beta \gamma \delta} \!
			\left(
				\BoldVec{k}, \BoldVec{p} , \BoldVec{q}
			\right)
			\langle
				\Annihilation{c}{
						\BoldVec{p}+\BoldVec{q} 3 \gamma 
				}
				\Annihilation{c}{
						\BoldVec{k}-\BoldVec{q} 4 \delta
				}
			\rangle
			\\ \nonumber
			g^{\mu}_{\alpha \delta} \! 
			\left( 
				\BoldVec{k},\BoldVec{q}
			\right)
			&
			=
			\frac{1}{2} 
			\sum_{
					\BoldVec{p}							
					\beta \gamma						
					1234															
			}
			[\sigma^\mu]_{14} 
			V^{1234}_{\alpha \beta \gamma \delta} \!
			\left(
				\BoldVec{k}, \BoldVec{p} , \BoldVec{q}
			\right)
			\langle
				\Creation{c}{
						\BoldVec{p} 2 \beta 
				}
				\Annihilation{c}{
						\BoldVec{p}+\BoldVec{q} 3 \gamma 
				}
			\rangle.
		\end{align}
		Here, $\sigma^\mu$ are the Pauli matrices including the identity and $\chi^\mu = \sigma^\mu (i \sigma^y)$. The first components of both $d^{\mu}$ and $g^{\mu}$ behave as scalars under spin rotations and correspond to spin-singlet superconductivity and charge orders, respectively. The last three components transform as vectors and thus correspond to spin-triplet superconductivity and magnetic orders, respectively.
		
		An ordered state with a symmetry that is incompatible with the symmetry of the normal state $H_0$ can obtain a finite value only after a spontaneous symmetry breaking at some critical temperature. Since $H_0$ is both spin and particle conserving, all PP orders are necessarily symmetry breaking. Many PH orders also break at least one symmetry, such as translational invariance for charge-density-waves or spin-rotation symmetry for magnetic states. Symmetry conserving PH orders can however be finite at any temperature and simply renormalize the band structure of $H_0$. Since we are considering competing orders with similar critical temperatures, we can safely assume that all band renormalizations are temperature independent and already included in $H_0$.
	
		A symmetry breaking order parameter is necessarily vanishingly small near its critical temperature, and $H_{\rm MF}$ is therefore only a small perturbation to the normal state. We can therefore calculate (see Appendix \ref{APP:selfC} for details) the response of the system and evaluate the expectation values in Eq.~\eqref{eq:DandG} from the first order perturbation of $H_{\rm MF}$ to the statistical ensemble density matrix. The result is a set of self-consistency equations that both have the same symmetry as $H_0$ and that are linear in the order parameters $g^{\mu}$ and $d^{\mu}$, which therefore do not mix. With the order parameters gathered in vectors $D\PPorPH$ with $+(-)$ superscript for the PP(PH) channel, the self-consistency equations have the form,
		\begin{equation} \label{eq:SelfC}
			D\PPorPH = \beta \mathbb{V}\PPorPH \mathbb{W}\PPorPH D\PPorPH
			,
		\end{equation}
		where all the interactions enter through the matrices $\mathbb{V}\PPorPH$, with different contributions to the two channels, as indicated by the superscript. The temperature enters both through an explicit factor of $\beta = 1/T$ and through the diagonal polarizability matrices $\mathbb{W}\PPorPH$ with the elements,
		\begin{equation} \label{eq:defW}
				W \PPorPH \! \left( \beta \xi_1 , \beta \xi_2 \right)
				=
				\frac{
					\tanh \left( \frac{ \beta \xi_1 }{2} \right) \pm \tanh \left( \frac{ \beta \xi_2 }{2} \right)
				}{
					\beta (\xi_1 \pm \xi_2 ) / 2
				}
				,
			\end{equation}
		where $\xi_1$ and $\xi_2$ are the energies of the interacting quasiparticles that enter the expectation values in Eq.~\eqref{eq:DandG}. The response of the order parameters to the perturbation $H_{\rm MF}$ are given by $\beta \mathbb{V}\PPorPH \mathbb{W} \PPorPH D\PPorPH$, where then the stability of the system is determined by the response matrices $\beta \mathbb{V}\PPorPH \mathbb{W}\PPorPH$. The eigenvectors of these matrices represents the possible orders, whose stability are given by the eigenvalues. An order whose eigenvalue is larger than 1 is an instability of the system, which is amplified by the response. If the eigenvalue is instead smaller than 1, then the order does not represent an instability but it decays through the response. All responses are usually small at high temperatures because of the explicit $\beta$-factor, but upon cooling the eigenvalues tend to grow. If at a critical temperature an eigenvalue grows to 1, then that order becomes the physical instability of the system and Eq.~\eqref{eq:SelfC} is satisfied also for a non-trivial zero solution. Thus, the critical temperatures of all possible orders are implicit in Eq.~\eqref{eq:SelfC}, with the leading order being the first instability appearing on cooling. 
		
		The general form of Eq.~\eqref{eq:SelfC} determines the behavior of all orders for a general set of interaction terms. However, this equation is much simpler for specific orders and interactions. For a pair scattering potential $V_{\BoldVec{k}\BoldVec{k}'}$ generating prototypical BCS pairing (spin-singlet pairing with opposite momentum states in a single band system), Eq.~\eqref{eq:SelfC} directly gives the linearized BCS gap equation,
		$	
			\Delta_{\BoldVec{k}} = 
			\sum_{\BoldVec{k}'} 
			2
			V_{\BoldVec{k} \BoldVec{k}'}
			\tanh ( \beta \xi_{\BoldVec{k}'} / 2)
			\Delta_{\BoldVec{k}'}
			/ (\beta \xi_{\BoldVec{k}'})
		$, since 
		$
			W \PP \! 
				\left(
					\beta
					\xi_{\BoldVec{k}}
					,
					\beta
					\xi_{ - \BoldVec{k}}
				\right)
			=
			2 \tanh ( \beta \xi_{\BoldVec{k}} / 2) / \beta \xi_{\BoldVec{k}}
		$, when $\xi_{\BoldVec{k}} = \xi_{-\BoldVec{k}}$. 
		Furthermore, in the PH channel for a magnetic exchange interaction $U$, an eigenvalue growing towards 1 for Eq.~\eqref{eq:SelfC} in the zero-temperature limit directly generates the the Stoner criterion for ferromagnetism: $U \rho(\epsilon_F) \geq 1$, with $\rho(\epsilon_F)$ the density of states at the Fermi level. Note that since the interactions contribute differently to the two PH and PP channels as well as the different orders, the interactions can drive many different instabilities simultaneously. For example, in a material an exchange interaction might drive a magnetic instability, while simultaneously a competing superconducting instability is driven by an effective electron-phonon interaction. Alternatively, the same interaction can also drive instabilities in both channels. In Section \ref{sec:ex} we provide realistic examples of both of these cases.
		\begin{figure*}[tbh]
			\includegraphics[width=1\textwidth]{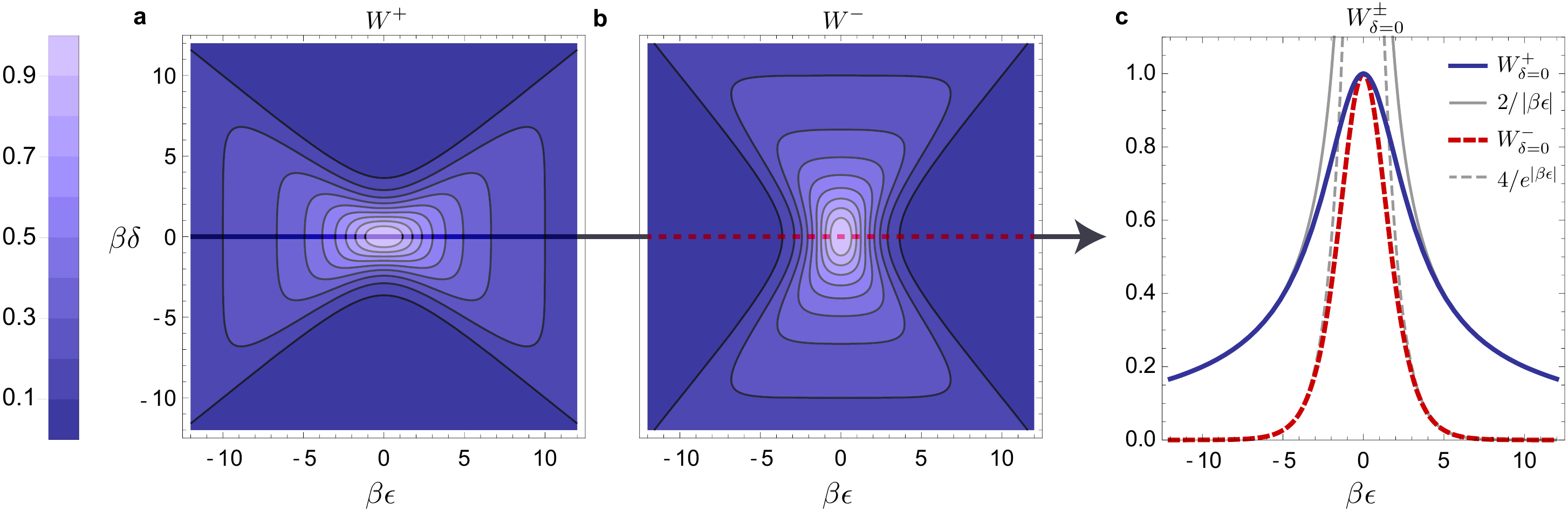}
			\caption{
				{(Color) Channel dependent polarizability.} 
				(\textbf{a}) the polarizability of the particle-particle (PP) channel, $W\PP$, and (\textbf{b}) the particle-hole (PH) channel, $W\PH$, as functions of the average energy of two interacting quasiparticles relative to the Fermi level, $\epsilon = (\xi_1 + \xi_2)/2$, and energy difference, $\delta = (\xi_1 - \xi_2)/2$. (\textbf{c}) the polarizability for two quasiparticles at equal energies, i.e. the indicated $\delta = 0$ cut through $W\PPorPH$. The asymptotic forms (gray lines) show that the PH polarizability decays exponentially with $\epsilon$, while the PP polarizability decays algebraically with $\epsilon$.
			}
			\label{fig:weight}
		\end{figure*}
		
\subsection{Channel dependent polarizability}
		Different orders have their broken symmetries and critical temperatures determined by the interactions in $\mathbb{V}\PPorPH$, but also by the ability of the quasiparticles to respond to these interactions, which is determined by the polarizabilities $ W\PPorPH $. We plot $ W\PPorPH $ in Fig.~\ref{fig:weight} as function of the average energy of the two quasiparticles relative to the Fermi level, $\epsilon = (\xi_1 + \xi_2) / 2$, and their energy difference, $\delta = (\xi_1 - \xi_2) / 2$. If both are small, then both quasiparticles are near the Fermi level and in both channels their polarizability attains the same maximal value. However, as the quasiparticle energies stray away from the Fermi level, the polarizabilities decrease in disparate ways. In Fig.~\ref{fig:weight}(c), the polarizability of two equal energy quasiparticles, i.e~the $ \delta =0 $ cuts through $W\PPorPH$, decay exponentially with $\epsilon$ in the PH channel but merely algebraically in the PP channel, through the asymptotic forms $W\PP_{\delta=0} \sim 2 \Abs{ \beta \epsilon }^{-1} $ and $ W\PH_{\delta=0} \sim 4 \exp ( - \Abs{ \beta \epsilon } )$. Clearly, if the average energy of two quasiparticles is larger than their energy difference, then they are better able to contribute strongly to PP orders than to PH orders. Thus PP orders benefit from states that are close together but potentially far from the Fermi level, while the opposite is true for PH orders. This is exactly the case for flat band systems.				
		
\subsection{Critical temperatures of flat bands}
		The response of flat bands and similar DOS peaks readily overshadow all other contributions to the order response when they are near the Fermi level. Further, the narrow width of the DOS peak limits the energy difference of interacting quasiparticles. Therefore the polarizabilities are uniform over the peak states and they only depend on the position of the DOS peak relative to the Fermi energy. Eq.~\eqref{eq:SelfC} therefore simplifies to $D\PPorPH = \beta W_{\delta=0} \PPorPH \mathbb{V}\PPorPH D\PPorPH$ for flat band systems. Further, when the DOS peak and the Fermi level exactly align, i.e. then $\epsilon=0$ and $\delta=0$, we find $W\PPorPH=1$. Thus, if $\nu\PPorPH$ is an eigenvalue of $\mathbb{V}\PPorPH$, then $\hat{T}_c\PPorPH = \nu\PPorPH$, and therefore the critical temperatures of both the PP and PH channels are directly proportional to the interaction strength. This linear relationship is an unusually strong dependence on the interaction strength, and therefore flat bands readily have very large critical temperatures.
		
		\begin{figure*}[t]	
			\includegraphics[width=1\textwidth]{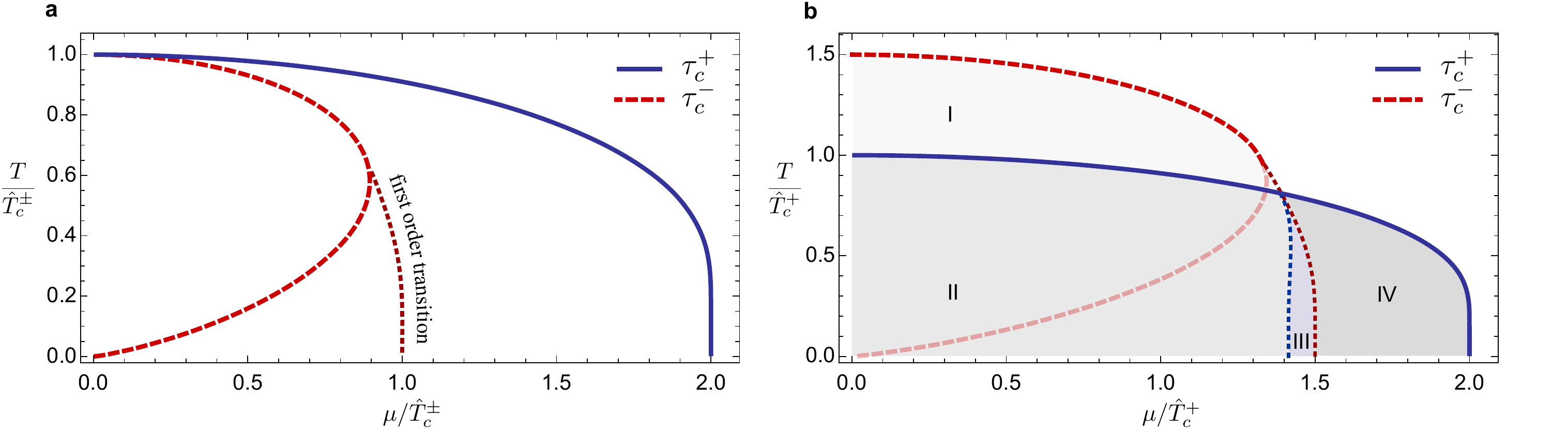}
			\caption{
				{ (Color) Universal phase diagrams for flat band systems.} 
				(\textbf{a}) the scaled critical temperatures of both PP orders $\tau_c\PP$ and of PH orders $\tau_c\PH$ as function of energy (doping) away from a flat band. The critical temperature and the doping $\mu$ are both scaled by the critical temperature $\hat{T}_c\PPorPH$ that results when both the Fermi level and the flat band align. The critical temperatures decrease according to Eqs.~\eqref{eq:ScaledTc} differently for the two channels, with the superconducting PP channel being more resilient to doping. PH orders also have a first order transition line to the normal state as shown. (\textbf{b}) revealing explicit example for when a PH order initially dominates over a PP order by $\hat{T}_c\PH = 1.5 \hat{T}_c\PP$ (see Appendix \ref{APP:peak} for details). The critical temperatures and the doping away from the flat band are here both scaled by $\hat{T}_c\PP$. There are four regions of interest, labeled by roman numerals and described in the text. Even though the PH order is initially much stronger, regions III and IV form a clear superconducting dome.
			}
			\label{fig:Phase}
		\end{figure*}				
		When the Fermi level is tuned away from the DOS peak, the critical temperatures decrease. If $\mu$ is the energy difference between the DOS peak and the Fermi level, then the critical temperatures of the PP channel $T_c\PP = 1 / \beta_c\PP$ and of the PH channel $T_c\PH = 1 / \beta_c\PH $ satisfy, respectively, $\tanhc \! \left( \beta_c\PP \mu / 2 \right) \beta_c\PP \nu\PP = 1$ and $\text{sech}^2 \! \left( \beta_c\PH \mu / 2 \right) \beta_c\PH \nu\PH = 1$. Further, the eigenvalues $\nu\PPorPH$ are constant if we assume that the interactions are unaffected by moving the Fermi level over the energy scale of $\hat{T}_c\PPorPH$. This is true for Coulomb interactions as well as for the effective electron-phonon BCS interaction, since it is attractive for $ \Abs{ 2 \delta} = \Abs{(\xi_k - \xi_l)} < \hbar\omega_D $ \cite{Tinkhambook}. With this assumption, the critical temperatures of both channels each follow their own unique universal curve when the Fermi level is tuned away from the DOS peak, with the overall energy scale set by $\hat{T}_c\PPorPH$ which encapsulates all interaction details. Using the dimensionless variables $\tau_c\PPorPH = T_c\PPorPH / \hat{T}_c\PPorPH$ and $\mu\PPorPH = \mu / \hat{T}_c\PPorPH$ we arrive at
			\begin{align}
				\tau_c\PP
				=
				\tanhc \! \left(
					\frac{ \mu\PP }{ 2 \tau_c\PP }
				\right)
				\eqText{and}
				\tau_c\PH
				=
				\text{sech}^2 \! \left( 
					\frac{ \mu\PH }{ 2 \tau_c\PH }
				\right) 
				\label{eq:ScaledTc}
				.			
			\end{align}
			
		Eq.~\eqref{eq:ScaledTc} is exact for a single flat band. It applies further to DOS peaks that are sufficiently narrow and that have a large order response contribution compared to the background states outside the peak. A peak is sufficiently narrow if its width $ \Delta E $ is narrow on the scale of the interaction strength, $ \Delta E \ll \nu_\text{max}\PPorPH / e$, where $\nu_\text{max}\PPorPH$ is the largest eigenvalue of $\mathbb{V}\PPorPH$ in Eq.~\eqref{eq:SelfC} (see Appendix \ref{APP:peak} for details). The peak and background states partition the response matrices, and they mix through the off-diagonal blocks. Thus, the background can influence the overall response, but it can only increase the critical temperature of the leading order of the peak due to Cauchy's interlace theorem \cite{horn2012matrix}. The influence is small if either the response of the background or the strength of the mixing is weak, i.e.~if the background DOS within the width of $W\PPorPH$ is small relative to the peak DOS or if the interactions between the peak and the background are small. Since the width of $W\PPorPH$ increases with the temperature, small deviations from the universal equations~\eqref{eq:ScaledTc} can start to appear at higher temperatures due to mixing with background states.

\subsection{Universal phase diagrams of flat bands}		
		We plot the solutions to Eqs.~\eqref{eq:ScaledTc} in Fig.~\ref{fig:Phase}(a) as function of the scaled doping $\mu\PPorPH$ away from the flat band. The PH order curve has two branches. The upper branch marks the onset of the PH order instability, which is an order transition that lowers the free energy. However, the lower branch does not mark an order transition. While it does mark a solution to Eq.~\eqref{eq:SelfC}, and as such the vanishing of the first derivative of the free energy with respect to the order parameter, it is not a viable transition because higher order terms increase the free energy. Because the upper branch does not enclose a region it has to end, as indicated in a first order transition line. This is inferred from Eq.~\eqref{eq:ScaledTc}, even though the transition line is not a solution to it. The shape of the line may therefore be influenced by the characteristics of both the interactions and the band structure, but for definiteness we plot the transition line for a representative ferromagnetic order (see Appendix \ref{APP:PD} for details). The line starts from where the two branches meet at $\mu\PH \approx 0.9$ and $\tau_c\PH \approx 0.6$, and ends at $\mu\PH = 1$ and $\tau_c\PH = 0$. Thus, no PH order survives beyond this doping region, but superconducting PP orders extend all the way out to $\mu\PP = 2$, due to the long-ranged polarizability of the PP channel in Fig.~\ref{fig:weight}(c). Thus, if a superconducting state is initially stronger than all PH states, $\hat{T}_c\PP > \hat{T}_c\PH$, then superconductivity will be favored for all doping levels, and the phase diagram has a superconducting dome firmly centered on the flat band. But even if a PH state is initially stronger, $\hat{T}_c\PP < \hat{T}_c\PH$, two superconducting domes still appear next to the flat band, since for $\hat{T}_c\PH \lesssim 2 \hat{T}_c\PP$ all PH orders end before the superconducting order, which leaves it uncontested on the flanks.
				
		In Fig.~\ref{fig:Phase}.(b), we show the example $\hat{T}_c\PH = 1.5 \hat{T}_c\PP$, which we for definiteness have calculated using a conventional $s$-wave superconducting PP order and a ferromagnetic PH order (see Appendix \ref{APP:PD} for details). In region I the PH order is alone viable, and so is the PP order in region IV. But both orders overlap in region II and III. In region II the PH order dominates in this specific example and also under most ordinary circumstances that exclude order mixing or crossing, since it is established first upon cooling.
		Region III is bounded by two lines. The outer line is the first order transition for the PH order. At this line, the free energy of the PH order and the normal state are equal, but the free energy of the PP order is generally decisively lower than both of these, since it is well established in this region. The superconducting PP order therefore extends past the PH first order line to an inner line where instead the free energies of the PP and PH orders cross. Together regions III and IV comprise a superconducting dome on the flanks of the flat band. 				
		
\section{FLAT BAND BEHAVIOR IN ABC-GRAPHITE AND DOPED GRAPHENE}
\label{sec:ex}		
		\begin{figure*}[t]
			\includegraphics[width=0.99\textwidth]{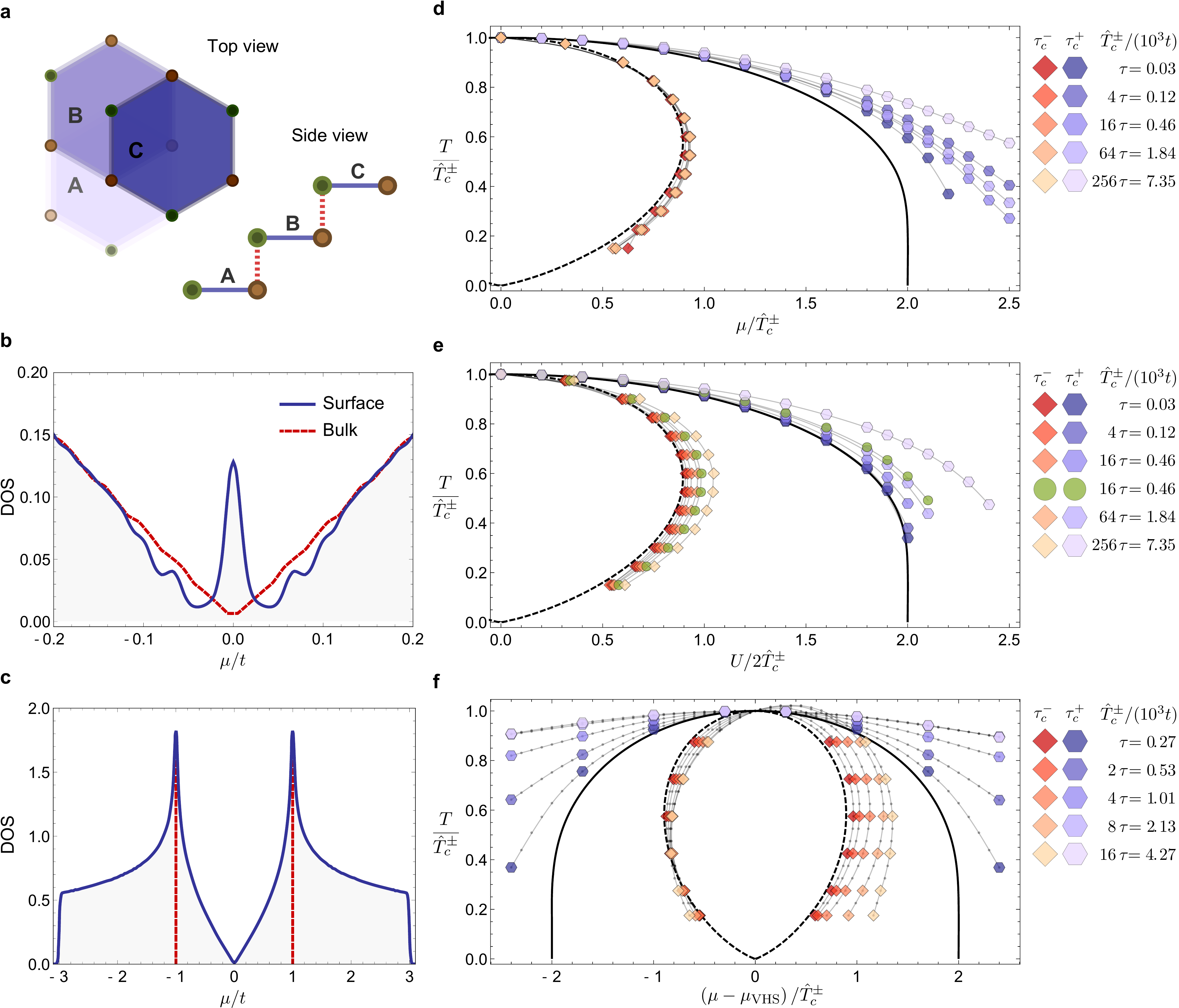}
			\caption{
				{ (Color) Universal critical temperature curves for finite ABC-stacks of graphite and heavily doped graphene.} (\textbf{a}) top and side view of the graphene sheet stacking of ABC-stacked graphite. (\textbf{b}) Bulk and surface DOS for ABC-stacked graphite in an 8 layer ABC-stacked graphite slab. (\textbf{c}) graphene DOS with VHS marked at $\mu = \pm t$. (\textbf{d}) and (\textbf{e}) normalized critical temperatures for surface ferrimagnetic (diamonds) and $s$-wave superconductivity orders (hexagons) in 8 layers of ABC-stacked graphite as a function of doping $\mu$ (\textbf{d}) and electric potential difference $U$ between top and bottom layer (\textbf{e}). The main data sets have an assumed linear electric potential profile across the graphite stack, while for the the green circle data set we used a non-linear profile with larger electric fields near the surfaces. (\textbf{f}) the normalized critical temperatures for SDW (diamonds) and chiral $d$-wave superconductivity (hexagons) orders as functions of doping $\mu$ away from the VHS in graphene. In (\textbf{d})--(\textbf{f}), the ideal flat band critical temperature curves from Eqs.~\eqref{eq:ScaledTc} are shown as black dashed (PH) and black solid (PP) curves. The maximum critical temperature $\hat{T}_c\PPorPH$ increases exponentially between each data set and for each data set polygons represent data points connected by thin lines to guide the eye, except for (\textbf{f}) where most data points have been suppressed to gray dots for clarity. 
			}
			\label{fig:systems}
		\end{figure*}
		After having derived the general behavior for flat band systems we now show explicitly how they apply in two real systems: rhombohedral ABC-stacked graphite with its topologically protected surface flat bands and heavily doped graphene with its logarithmically divergent VHS peaks. Graphene has a honeycomb lattice with a nearest-neighbor (NN) hopping $t$ \cite{RevModPhys.81.109}. Pristine graphene has two Dirac cones with a Fermi velocity $v_F$ at the $\pm K$ points. At the high doping $\mu = \pm t$, the Fermi surface intersects the $M$ point where the energy bands have locally flat saddle points that result in the divergent VHS DOS peaks seen in Fig.~\ref{fig:systems}(c). Graphene layers stacked in a repeating staircase fashion, as shown in Fig.~\ref{fig:systems}.(a), makes ABC-stacked graphite. An alternative stacking, AB-stacked Bernal graphite, is most common in nature, but the ABC stacking has been found in both graphite and multi-layer graphene \cite{PhysRevB.81.161410,pierucci2015evidence,henni2016rhombohedral}, and it is more stable in an electric field \cite{PhysRevB.84.233404}. With a NN interlayer hopping amplitude $t_\bot$, ABC-graphite is described by a Hamiltonian, 		
		\begin{align*}
			H_{\rm ABC}
			=	
			&
			- \! \sum_{\NN{i}{j},l,\sigma}{
					\left(
						t 
						\Hopping{a}{i l \sigma}{b}{j l \sigma}
						+
						t_{\perp}
						\Hopping{a}{i (l+1) \sigma}{b}{j l \sigma}
					\right)
					+ \text{H.c} 
					}	
			\\
			&
			- \mu \sum_{i,l,\sigma}{
					\left(
						\NumberOp{a}{i l \sigma}
						+
						\NumberOp{b}{i l \sigma}
					\right)
					}
			,
		\end{align*} 
		where $a^\dagger_{il\sigma}$ ($b^\dagger_{il\sigma}$) creates an electron in sublattice $A$ ($B$), in unit cell $i$, and in layer $l$, with the spin $\sigma$. Here $t \approx 3 \: \text{eV}$, $t_{\bot} \approx 0.4 \: \text{eV}$, and $\mu$ is the chemical potential \cite{PhysRevB.87.140503}. The intralayer hopping turns the graphene Dirac cones into Fermi spirals that carry a topological number. This results in topologically protected zero-energy surface flat band states for $\Abs{\BoldVec{q}} < t_\bot / v_F$ on surfaces perpendicular to the stacking direction, where $\BoldVec{q}$ is the momentum measured from $\pm K$ \cite{PhysRevB.90.085312,FBTM1,FBTM2,PhysRevB.84.165404}. The resulting large surface DOS peak is seen at the center of the linearly vanishing bulk DOS in Fig.~\ref{fig:systems}(b). The topological protection is exact when the lattice has sublattice symmetry. Higher order intra-sublattice site hopping terms lifts this symmetry and the protection. However this does not introduce any qualitative changes to our results, as the small additional hopping terms do not significantly affect the large DOS peak \cite{PhysRevB.87.140503,PhysRevB.84.165404}.
		
		The large surface DOS peak is very susceptible to develop a finite order, and ab-initio calculations find a strong ferrimagnetic ordering \cite{PhysRevB.81.161403,Cuong2012253}. We capture this order by accounting for interaction with a repulsive Hubbard-$U$ term $H_{\rm U} = U \sum_{i} n_{i\uparrow} n_{i\downarrow}$ for each site $i$. Solving self-consistently for the PH magnetic order parameters, we find a stable collinear ferrimagnet, with unequal magnetic moments between the two sublattices, in agreement with the ab-initio calculations. The sublattice asymmetry stems from the weight of the flat band surface state being concentrated to one of the sublattices for each surfaces of a slab. Similarly, assuming conventional $s$-wave superconductivity achieved by electron-phonon interactions, we find a superconducting state using $H_{\rm SC} = - V \sum_{i}  n_{i\uparrow} n_{i\downarrow}$ with $V>0$ \cite{PhysRevB.83.220503}. Moreover, it has been shown that an infinite graphite stack acquires a gapless fluctuation mode that readjusts the mean-field results \cite{PhysRevB.93.024505}. The gap is however finite for all finite-sized stacks, and mean-field theory is therefore valid for the finite stacks considered here.
				
		In Fig.~\ref{fig:systems}(d) we show, as functions of a uniform doping away from the surface flat band, the normalized critical temperatures of both the ferrimagnetic and the superconducting order. Both orders follow closely the ideal flat band prediction of Eqs.~\eqref{eq:ScaledTc} (black lines) for a wide range of coupling strengths and critical temperatures. For very large temperatures, the mixing with the background states causes a small increase in $T_c$, which sustains both orders further than the idealized flat band solution, but especially the superconducting order.
		The surface carrier occupation can also be adjusted by applying an electric field in the stacking direction, since the electric potential both acts, and can be modeled, as a layer dependent chemical potential, $\mu_l$. Fig.~\ref{fig:systems}(e) shows that, as a function of the potential difference between the top and bottom surface $U$, the critical temperatures of both PH and PP orders again follow closely the ideal flat band prediction of Eqs.~\eqref{eq:ScaledTc}. Moreover, we find this result largely independent of the potential profile across the graphite stack. Allowing $\mu_l$ to be an odd polynomial centered around the middle of the stack, we plot in Fig.~\ref{fig:systems}(e) both the linear (red/blue) and the third order (green) electric potential profiles, where higher order gives a steeper (screened) potential profile. Thus, even if ABC-stacked graphite is initially found in a magnetic state, simply applying an electric field can reveal an underlying superconducting state.
				
		Finally we study heavily doped graphene around the van Hove singularity. Here recent (f)RG results have found a very close competition between chiral $d$-wave superconducting and SDW orders \cite{nandkishore2012chiral,PhysRevB.86.020507,PhysRevB.85.035414}. At the mean-field level we can capture both these orders with a single antiferromagnetic NN Heisenberg spin interaction
		$
		H_{\rm J} = -J \sum_{\NN{i}{j}}{
			\left(
				\BoldVec{S}_i \cdot \BoldVec{S}_i - n_i n_j /4 
			\right)
		}
		$. The hexagonal symmetry automatically favors the chiral $(d \pm i d')$-wave in the PP channel for this interaction \cite{0953-8984-26-42-423201,ABSPhysRevB.75.134512,TLandABSPhysRevB.90.224504}. For PH orders we find using an eight site supercell, the very recently proposed uniaxial SDW of Ref.~\cite{PhysRevLett.108.227204}. Even though the bands at the VHS are only locally flat, as opposed to regionally flat, we nonetheless find that the critical temperatures in Fig.~\ref{fig:systems}(f) also closely follow Eqs.~\eqref{eq:ScaledTc} for a wide range of coupling strengths and critical temperatures for both the SDW and the chiral $(d \pm i d')$-wave superconducting state.
		Deviations from the ideal behavior in Fig.~\ref{fig:systems}(f) are also readily explained, since they follow from the overall DOS, see Fig.~\ref{fig:systems}(c), and the polarizabilities. Because the DOS is larger on the outer side of the VHS and because PH orders benefit from interacting states with different energies, the PH order is more strongly affected by the background compared to the PP order. Therefore the PH order has its maximal critical temperature slightly above the VHS as a result. Thus, our general flat band results dictate that even if a SDW is established at the VHS, chiral superconductivity domes may still exist on the flanks of the VHS, with the dome on the inner side somewhat larger than that on the outer side. Exactly such a characteristic superconducting dome structure has been found by recent fRG calculations \cite{PhysRevB.86.020507,PhysRevB.85.035414}. 
		
\section{CONCLUSIONS}		
In summary we have shown that electronic systems with flat, or nearly flat, energy dispersions are a very fertile ground for all types of ordered states. In fact, the strong DOS peaks give rise to a phase diagram where all PH and PP orders follow their own universal curves for the critical temperature as function of doping away from the flat band. We find that both PH and PP critical temperatures scale linearly with interaction strength when the flat band coincides with the Fermi level. At finite doping, however, the PP orders are significantly more resilient and survives to much larger doping levels. Thus, even if magnetic or charge order dominated initially, superconducting domes are likely to appear on the flanks of flat bands, accessible through doping or tuning with electric fields. Since our results are only relying on a strong DOS peak close to the Fermi level, the results also apply to systems with van Hove singularities or similar large DOS peaks. To illustrate the applicability of our results we also provide two case studies. For both the flat surface bands of rhombohedral ABC-stacked graphite and the van Hove singularity in heavily doped graphene we find superconducting domes appearing with doping and/or electric field, even when magnetism initially dominates. This clearly illustrates how flat band systems offer a tantalizing route towards realizing high-temperature superconductivity. 

\section{ACKNOWLEDGEMENTS}
This work was supported by the Knut and Alice Wallenberg Foundation (KAW), the Swedish Research Council (Vetenskapsr\aa det), the G\"{o}ran Gustafsson Foundation, and the Swedish Foundation for Strategic Research (SSF).

\appendix

\section{DERIVATION OF SELF-CONSISTENCY EQUATIONS} \label{APP:selfC}
	The self-consistency equations~\eqref{eq:SelfC} are derived by perturbation theory in the density matrix $\hat{\rho} = \exp ( - \beta \hat{H} )$. Since  $\partial \hat{\rho}/\partial \beta = - \hat{H}\hat{\rho}$, we find $\frac{\partial}{\partial \beta} ( \Eexp{ \beta \hat{H}_0 } \hat{\rho}) = - \Eexp{ \beta \hat{H}_0} \hat{H}_{\rm MF} \hat{\rho}$ \cite{Feynman199803}, where this latter equation has the formal solution,
	\begin{align*}
			\hat{\rho}
			=
			\hat{\rho}_0
			\mathcal{T}
			\Eexp{
				- \int_0^\beta \dd \beta' 
				H \! \left( \beta' \right) 
			}
			=
			\hat{\rho}_0
			-
			\hat{\rho}_0
			\int_0^\beta \dd \beta' 
			H \! \left( \beta' \right) 
			+
			\cdots,
		\end{align*}
	where $\mathcal{T}$ is the time-ordering operator and $H (\beta) = \Eexp{\beta H_0}H_{\rm MF}\Eexp{-\beta H_0}$. Thus, to first order in $H_{\rm MF}$, $\Delta \hat{\rho} = - \hat{\rho}_0 \int_0^\beta \dd \beta' H(\beta)$, which is the response of the system to infinitely small order parameters. Using $H_0$ in Eq.~\eqref{eq:H}, we find
		\begin{align*}
			&
		\Delta \hat{\rho} =
		-
		\hat{\rho}_0
		\sum_{
			{
				\BoldVec{k} \BoldVec{p}	
				\alpha \beta						
				12															
			}
		}
		\left[
			\frac{
				\Eexp{
					\beta 
					\left[
						\xi_{\alpha} \! \left( \BoldVec{k} \right)
						+
						\xi_{\beta} \! \left( \BoldVec{p} \right)
					\right]
				}
				-
				1			
				}{
					\xi_{\alpha} \! \left( \BoldVec{k} \right)
					+
					\xi_{\beta} \! \left( \BoldVec{p} \right)
			}
			\right. \times
			\\
			&
			\left.
			\qquad \qquad \qquad \qquad
			\left[
				d_{\alpha \beta} \! 
				\left( 
					\BoldVec{k},\BoldVec{p}
				\right)
				\cdot
				\chi
			\right]_{12}
			\Creation{c}{
				{
					\BoldVec{k}		
					1 \alpha
				}
			}
			\Creation{c}{
				{
					\BoldVec{p}	
					2 \beta
				}
			}
			\right]
			+
			\text{H.c.}
		\\
		&
		- 
		4
		\hat{\rho}_0
		\sum_{
			{
				\BoldVec{k} \BoldVec{q}	
				\alpha \delta						
				14															
			}
		}
		\frac{
			\Eexp{
				\beta 
				\left[
					\xi_{\alpha} \! \left( \BoldVec{k} \right)
					-
					\xi_{\delta} \! \left( \BoldVec{p} \right)
				\right]
			}
			-
			1			
			}{
				\xi_{\alpha} \! \left( \BoldVec{k} \right)
				-
				\xi_{\delta} \! \left( \BoldVec{p} \right)
		}
		\left[
			g_{\alpha \delta} \! 
			\left( 
				\BoldVec{k},\BoldVec{q}
			\right)
			\cdot
			\sigma
		\right]_{14}
		\Creation{c}{
			{
				\BoldVec{k}			
				1 \alpha
			}
		}
		\Annihilation{c}{
			{
				\BoldVec{k}-\BoldVec{q}		
				4 \delta
			}
		}.
	\end{align*}
	
	We then evaluate the expectation values entering the definitions of the PP and PH order parameters in Eq.~\eqref{eq:DandG} using the density matrix $\hat{\rho} = \hat{\rho_0} + \Delta\hat{\rho}$. This gives a two particle expectation expression, which because of Wick's theorem is the product of two Fermi functions dependent on the two quasiparticle energies $\xi_1$ and $\xi_2$. The result is the self-consistent equations~\eqref{eq:SelfC}, which have the full form		
	\begin{align*} 
			&
			d^{\mu}_{\alpha \beta} \! 
			\left( 
				\BoldVec{k},\BoldVec{p}
			\right)
			=
			\frac{1}{4}
			\sum_{
				{
					\BoldVec{q}							
					\gamma \delta						
					1234															
				}
			}
			\beta
			V^{1234}_{\alpha \beta \gamma \delta} \!
			\left(
				\BoldVec{k}, \BoldVec{p} , \BoldVec{q}
			\right) 
			[\chi^{\mu}]^\dagger_{12}
			\times
			\nonumber \\
			&
			\quad
			W\PP \! 
				\Big(
					\beta
					\xi_{\delta} \! \left( 
						\BoldVec{k} - \BoldVec{q}
					\right)
					,
					\beta
					\xi_{\gamma} \! \left( 
						\BoldVec{p} + \BoldVec{q}
					\right)
				\Big)
			\left[
				d_{\delta \gamma} \! 
				\left( 
					\BoldVec{k} - \BoldVec{q}, \BoldVec{p} + \BoldVec{q}
				\right)
				\cdot
				\chi
			\right]_{43}
			\nonumber \\
			&
			g^{\mu}_{\alpha \delta} \! 
			\left( 
				\BoldVec{k},\BoldVec{q}
			\right)
			=
			-
			\frac{1}{2}
			\sum_{
				{
					\BoldVec{p}							
					\beta \gamma					
					1234															
				}
			}
			\beta
			V^{1234}_{\alpha \beta \gamma \delta} \!
			\left(
				\BoldVec{k}, \BoldVec{p} , \BoldVec{q}
			\right) 
			[\sigma^{\mu}]_{41}
			\times
			\nonumber \\ 
			&
			\quad
			W\PH \! 
				\Big(
					\beta
					\xi_{\gamma} \! \left( 
						\BoldVec{p} + \BoldVec{q}
					\right)
					,
					\beta
					\xi_{\beta} \! \left( 
						\BoldVec{p}
					\right)
				\Big)
			\left[
				g_{\gamma \beta} \! 
				\left( 
					\BoldVec{p} + \BoldVec{q}, \BoldVec{q}
				\right)
				\cdot
				\sigma
			\right]_{32}
			.
		\end{align*}
		Schematically these equations are expressed in Eq.~\eqref{eq:SelfC}. Here all the dependence on the interacting quasiparticle energies $\xi_1$ and $\xi_2$, from both the perturbation and the occupation functions, are captured in the polarizability factors defined in Eq.~\eqref{eq:defW}.
		 
\section{PEAK WIDTH CONDITION} \label{APP:peak}
		Equation~\eqref{eq:ScaledTc} applies to DOS peaks that are sufficiently narrow to have an approximately uniform polarizability, which is true if its states obey $\beta \delta \ll 1$ near the critical temperature. Since $\delta$ is limited by the peak width $\Delta E$, the condition $\Delta E \ll T_c$ is sufficient. A lower bound for $T_c$ is given by $\nu_\text{max}\PPorPH w_\text{min}\PPorPH$ \cite{Lu2000}, where $\nu_\text{max}\PPorPH$ is the largest eigenvalue of $\mathbb{V}\PPorPH$ in Eq.~\eqref{eq:SelfC} and $w_\text{min}\PPorPH$ is the smallest polarizability among the peak states, which depends on $\Delta E$. Since the polarizability decays exponentially in the steepest direction, $\exp ( - \beta \Delta E ) < w_\text{min}\PPorPH$ when the Fermi level and the peak align. After solving for a lower bound on the critical temperature using this lower bound on $w_\text{min}$, the narrowness condition becomes $ \Delta E \ll \nu_\text{max}\PPorPH / e$. Thus, the peak has to be narrow on the scale of the interaction strength. 
		
\section{PHASE DIAGRAM IN FIG.~\ref{fig:Phase}} \label{APP:PD}
	Even though the general structure of the phase diagram in Fig.~\ref{fig:Phase} is given in Eqs.~\eqref{eq:ScaledTc}, we calculate for definiteness the phase diagram for a generic flat band system with a homogenous ferromagnetic PH order in competition with a conventional $s$-wave superconducting PP order. Letting $H_{\text{FM}}$ describe the PH order, driven by the interaction $J$, and $H_{\text{SC}}$ the PP order, driven by the interaction $V$, we determine the order parameter $m$ in
	$
		H_{\text{FM}} = 
		-
		\sum_{\BoldVec{k}}{
			\left(
				m
				[
					\NumberOp{c}{\BoldVec{k} \uparrow}
					-
					\NumberOp{c}{\BoldVec{k} \downarrow}
				]
				-
				m^2 / J
			\right)
		}
	$
	and $\Delta$ in 
	$
		H_{\text{SC}} = 
		-
		\sum_{\BoldVec{k}}{
			\left(
				[
					\Delta
					\BCSTermUp{c}{\BoldVec{k}}{c}{-\BoldVec{k}}
					+
					\text{H.c}
				]
				+
				\Abs{\Delta}^2 / V
			\right)
		}
	$
	by minimizing the free energies of $H_{\text{FB}} + H_{\text{FM}}$ and $H_{\text{FB}} + H_{\text{SC}}$, respectively, where
	$
		H_{\text{FB}} = \sum_{\BoldVec{k} \sigma} 
		(\epsilon(\BoldVec{k}) - \mu)
		\NumberOp{c}{\BoldVec{k} \sigma},
	$
	where $\epsilon(\BoldVec{k}) \rightarrow 0$ for all momenta, is the kinetic energy for a spin degenerate and completely flat system. Using the self-consistently calculated order parameters we evaluate the free energy for each ordered state and use them to find the four regions and free energy crossings in Fig.~\ref{fig:Phase}(b).
	

%

\end{document}